\documentclass[aps,prl,twocolumn]{revtex4}
\usepackage{amsmath}
\usepackage{graphicx}
\newcommand{\ltrsim}{\mathrel{\lower .3ex \rlap{$\sim$}\raise .5ex\hbox{$<$}}}
\begin{document}
\preprint{ }
\title{Broad Peak in the $d_{x^2-y^2}$ Superconducting Correlation Length as a
Function of Hole Concentration in the Two-Dimensional $t$-$J$ Model}
\author{W. O. Putikka}
\affiliation{Physics Department, The Ohio State University, 1680 University Dr., Mansfield,
OH 44906}
\author{M. U. Luchini}
\affiliation{31 Wingate Road, London W6 0UR, United Kingdom}

\begin{abstract}
We have calculated high temperature series to 12th order in inverse temperature for singlet
superconducting correlation functions of the 2D $t$-$J$ model with
$s$-, $d_{x^2-y^2}$- and $d_{xy}$-symmetry pairs.
We find the correlation length for $d_{x^2-y^2}$ pairing grows strongly with decreasing
temperature and develops a broad peak as a function of doping at $T/J=0.25$ for
$J/t=0.4$.  The correlation lengths for $s$- and $d_{xy}$-symmetry remain small and do
not display peaks.  Antiferromagnetic spin correlations at low doping act to suppress
the $d_{x^2-y^x}$ and $d_{xy}$ superconducting correlation lengths.
Our results support the hypothesis that the
strong electronic correlations found in the CuO$_2$ planes of high temperature
superconductors are the origin of the superconducting order.

\end{abstract}
\volumeyear{year}
\volumenumber{number}
\issuenumber{number}
\eid{identifier}
\date{\today}
\maketitle

Shortly after the discovery of high temperature superconductors by Bednorz and
M\"uller\cite{bednorz}, Anderson\cite{anderson} proposed that the new superconductors
should be viewed as single band doped Mott insulators, with a novel, non-phonon
origin for their superconductivity.  The model used to describe the CuO$_2$ planes
found in these materials is the two-dimensional Hubbard model, which in its
strongly correlated limit was restated by Zhang and Rice\cite{zhangrice} as the 2D $t$-$J$ model.
Below we show that the 2D $t$-$J$ model has strongly enhanced $d_{x^2-y^2}$
pairing correlations for physical choices of the model parameters and doping range.

The Hubbard and $t$-$J$ models have been widely studied by every method available in
theorists' arsenals\cite{threv}.  The results to date for pairing correlations have been at best
mixed\cite{hirsch, white, imada, scalapino, zhang, shih, sorella, pryadko, ogata}.  
Lacking small parameters,
strongly correlated models present severe difficulties for analytic calculations
and numerical techniques have difficulties with
finite size effects and the fermion sign problem.  
To avoid these problems, we calculate high temperature series
to 12th order in inverse temperature $\beta$ for the pairing correlations in the 2D
$t$-$J$ model.  The series are generated by a linked cluster expansion\cite{gelfand}, 
putting four extra operators on each cluster for the two pairs.

The Hamiltonian for the $t$-$J$ model is
\begin{equation}
H = -tP\sum_{\langle ij\rangle\sigma}\left(c_{i\sigma}^{\dagger}c_{j\sigma}+
c_{j\sigma}^{\dagger}c_{i\sigma}\right)P+J\sum_{\langle ij\rangle}{\bf S}_i\cdot{\bf S}_j,
\end{equation}
where the sums are over nearest neighbor pairs of sites on a square lattice and the $P$
operators eliminate states with doubly occupied sites from the
Hilbert space.  The spin operators are given by 
${\bf S}_i=\sum_{\alpha\beta}c_{i\alpha}
^{\dagger}\mbox{\boldmath $\sigma$}_{\alpha\beta} c_{i\beta}$, 
where $\mbox{\boldmath $\sigma$}_{\alpha\beta}$ is
a vector of Pauli matrices.

We limit our calculation to spin singlet pairing, with the pair operator for sites
$i$ and $j$ given by
\begin{equation}
\Delta_{ij} = \frac{1}{\sqrt{2}}\left(c_{i\uparrow}c_{j\downarrow}-c_{i\downarrow}c_{j\uparrow}\right),
\end{equation}
and the Hermitian adjoint
\begin{equation}
\Delta^{\dagger}_{ij} = \frac{1}{\sqrt{2}}\left(c_{i\downarrow}^{\dagger}c_{j\uparrow}^{\dagger}
- c_{i\uparrow}^{\dagger}c_{j\downarrow}^{\dagger}\right).
\end{equation}

The Hermitian real space four-point correlation function with operators on sites $i$, $j$, $k$
and $l$ is 
\begin{eqnarray}
P(i,j,k,l) & = & \frac{1}{2}\langle\Delta^{\dagger}_{ij}\Delta_{kl}+
                            \Delta^{\dagger}_{lk}\Delta_{ji}\rangle\\
           & - &\frac{1}{4}\left[
              \langle c_{i\downarrow}^{\dagger}c_{l\downarrow}\rangle\langle
               c_{j\uparrow}^{\dagger}c_{k\uparrow}\rangle+\langle c_{i\downarrow}^{\dagger}
               c_{k\downarrow}\rangle\langle c_{j\uparrow}^{\dagger}c_{l\uparrow}\rangle\right.
                   \nonumber \\
        &   &  +\langle c_{i\uparrow}^{\dagger}c_{l\uparrow}\rangle\langle c_{j\downarrow}
                   ^{\dagger}
              c_{k\downarrow}\rangle + \langle c_{i\uparrow}^{\dagger}c_{k\uparrow}\rangle
              \langle c_{j\downarrow}^{\dagger}c_{l\downarrow}\rangle\nonumber \\
    &   &  +\langle c_{l\downarrow}^{\dagger}c_{i\downarrow}\rangle\langle c_{k\uparrow}
            ^{\dagger}c_{j\uparrow}\rangle+\langle c_{k\uparrow}^{\dagger}c_{i\uparrow}\rangle
             \langle c_{l\downarrow}^{\dagger}c_{j\downarrow}\rangle\nonumber \\
   &   &  +\left.\langle c_{l\uparrow}^{\dagger}c_{i\uparrow}\rangle\langle c_{k\downarrow}^{\dagger}
          c_{j\downarrow}\rangle + \langle c_{k\downarrow}^{\dagger}c_{i\downarrow}\rangle
          \langle c_{l\uparrow}^{\dagger}c_{j\uparrow}\rangle\right],\nonumber
\end{eqnarray}
where we have removed the single particle contributions to $P(i,j,k,l)$ to focus on 
pairing fluctuations.  For a non-interacting system $P(i,j,k,l)=0$.
Due to the no double occupancy constraint in the $t$-$J$ model we have two restrictions on
the site choices: $i\ne j$ and $k\ne l$.  Otherwise the pairs are allowed to take on all
arrangements and sizes on the clusters.  

From our data we calculate series for the real space correlators $P({\bf r}, {\bf r}', {\bf R})$,
where {\bf r} and ${\bf r}'$ are vectors giving the orientation and internal size of the two
pairs and {\bf R} is the distance between the pairs, determined by the distance between the centers of mass.
Since the pairs are spin singlets,  {\bf r} and ${\bf r}'$ are limited to the upper
half plane and positive $x$-axis to avoid double counting.  All of the distances are measured in
terms of the lattice spacing $a$.  Since the original four point correlation function $P(i,j,k,l)$ is
identically zero for non-interacting systems, we also have $P({\bf r},{\bf r}',{\bf R})=0$ for non-interacting
systems.
The internal symmetry of the pairs is included by multiplying 
$P({\bf r}, {\bf r}', {\bf R})$ by phase factors $\phi_i({\bf r})$ and $\phi_i({{\bf r}'})$ where each
phase factor can have the values $+1$, $-1$ or $0$.  We have $\phi_i({\bf r})=0$ when {\bf r} is aligned
with a node.  Here the subscript $i$ labels the different symmetries we consider: $s$, 
$d_{x^2-y^2}$ or $d_{xy}$.
The phase factor of a single pair is shown in Fig. 1 for the anisotropic symmetries $d_{x^2-y^2}$
and $d_{xy}$.  An $s$-symmetry pair has uniform phase, $\phi_s({\bf r})=+1$ for all {\bf r}.

From $P({\bf r}, {\bf r}', {\bf R})$ and the phase factors
we want to construct a correlator $P_i({\bf R})$ which depends only
on {\bf R}, the separation between the pairs, and the symmetry of the pairs ($i=s$, $d_{x^2-y^2}$ or $d_{xy}$).  
This is then used to measure the strength of the pairing correlations by calculating
the ${\bf q}=0$ correlation length via
\begin{equation}
\xi_i^2 = \frac{1}{2d}\frac{\sum_{\bf R}|{\bf R}|^2 P_i({\bf R})}{\sum_{\bf R}P_i({\bf R})},
\end{equation}
with dimension $d=2$.

One way to calculate $P_i({\bf R})$ is to sum over {\bf r} and ${\bf r}'$ independently
\begin{equation}
P_i({\bf R})=\sum_{{\bf r}}\phi_i({\bf r})\sum_{{\bf r}'}\phi_i({{\bf r}'})P({\bf r}, {\bf r}', {\bf R}).
\end{equation}
We find that for the anisotropic pair states with $|{\bf R}|>0$ Eq. 6 gives $P_i({\bf R})$'s that are
very small and negative.  This suggests there is spurious cancellation occurring in our definition
of $P_i({\bf R})$.

As an example of the problem that can arise when calculating $P_i({\bf R})$ by Eq. 6, 
consider the special case\cite{conv} of pairs limited in size to nearest neighbors
with $d_{x^2-y^2}$ symmetry.  This is the type of calculation done in Refs. 
\onlinecite{hirsch, white, imada, scalapino, zhang, shih, sorella, pryadko} and Ref. \onlinecite{ogata}.
\begin{figure}[htb]
\includegraphics[width=3.4in,clip]{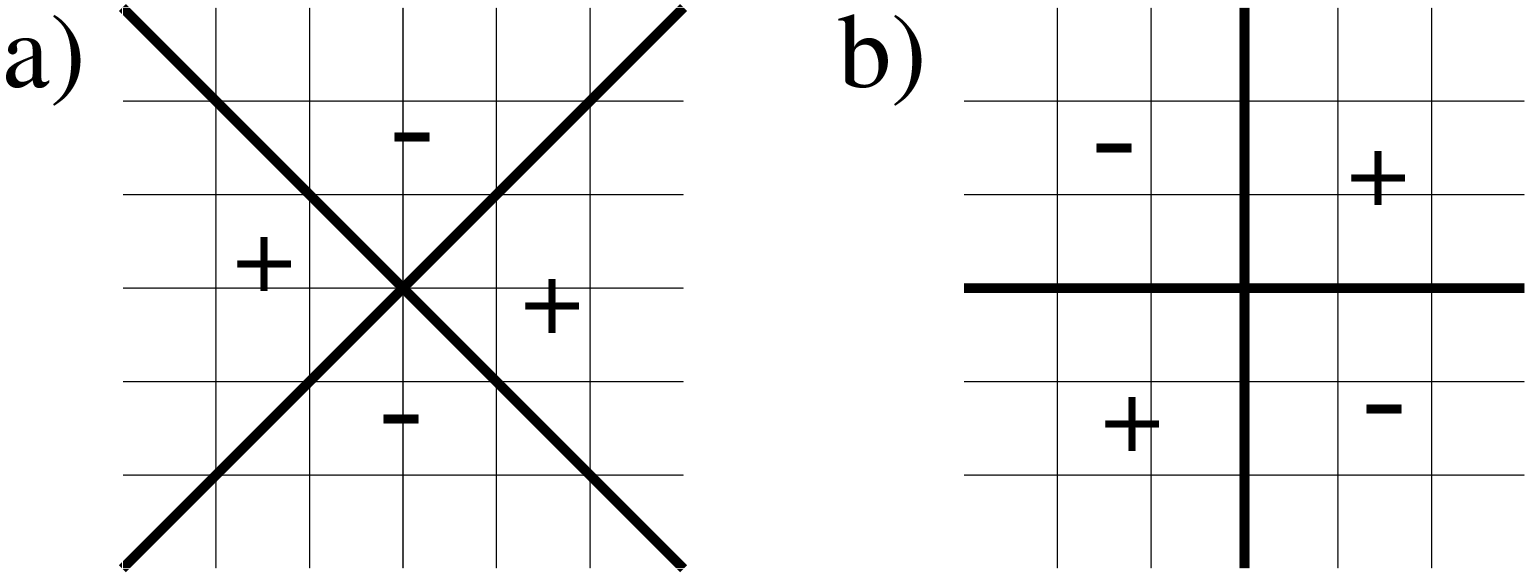}
\caption{Internal nodes and spatial phases for $d$-symmetry pairs: a) $d_{x^2-y^2}$,
b) $d_{xy}$.  The light 
lines are the underlying square lattice and the solid black lines are the nodal lines.
To determine the phase, consider one operator of a pair at the origin (where the nodes cross)
and put the second operator on a different lattice site.  The phase associated with that
part of the lattice is the internal phase of the pair.  The second operator cannot
be put on the nodal lines; the pair wave function is zero for those configurations.
For $s$-symmetry there are no nodes determined by
symmetry and the phase is always positive.}
\end{figure}
We have two types of correlators to consider (omitting the subtracted single particle terms
for clarity):
\[
\begin{array}[t]{lcr}
\langle O_+\rangle & = & \langle c_{i\downarrow}^{\dagger}c_{i+{\bf \hat{x}}\uparrow}
^{\dagger}c_{i+{\bf R}+{\bf \hat{x}}
\uparrow}c_{i+{\bf R}\downarrow}\rangle \\
   &   &   \\
\langle O_-\rangle & = & \langle c_{i\downarrow}^{\dagger}
c_{i+{\bf \hat{x}}\uparrow}
^{\dagger}c_{i+{\bf R}+{\bf \hat{y}}\uparrow}c_{i+{\bf R}\downarrow}\rangle,  
\end{array}
\]
where ${\bf\hat{x}}$ and ${\bf\hat{y}}$ are unit vectors in the $x$ and $y$ directions.  For
$d_{x^2-y^2}$ symmetry 
we have the phases $\phi_+=+1$ and $\phi_-=-1$.  Now 
consider what happens to $P_{x^2-y^2}({\bf R})$ above and below $T_{\rm c}$ as $|{\bf R}|\rightarrow\infty$:
\[
\begin{array}[t]{lclr}
\langle O_+\rangle & \rightarrow & |\Phi_0|^2 + A_<\exp(-|{\bf R}|/\xi_{x^2-y^2}) & T<T_{\rm c} \\
      &             &                                      &             \\
      & \rightarrow & A_>\exp(-|{\bf R}|/\xi_{x^2-y^2})              & T>T_{\rm c}
\end{array}
\]
and
\[
\begin{array}[t]{lclr}
\langle O_-\rangle & \rightarrow & -|\Phi_0|^2 + B_<\exp(-|{\bf R}|/\xi_{x^2-y^2}) & T<T_{\rm c} \\
      &             &                                        &             \\
      & \rightarrow & B_>\exp(-|{\bf R}|/\xi_{x^2-y^2})               & T>T_{\rm c},
\end{array}
\]
where $|\Phi_0|^2$ is the magnitude squared of the $d_{x^2-y^2}$ superconducting
condensate and the $A$'s and $B$'s are non-universal coefficients of the parts of
the correlator decaying with correlation length $\xi_{x^2-y^2}$.  Below $T_{\rm c}$ we then
have $P_{x^2-y^2}({\bf R})\sim \langle O_+\rangle-\langle O_-\rangle\sim |\Phi_0|^2$ as 
$|{\bf R}|\rightarrow\infty$ due
to the compensating phases for $\langle O_-\rangle$.  This is the definition of long range
superconducting order with $d_{x^2-y^2}$ symmetry.

High temperature series, however, are limited to the non-ordered high temperature phase where
$P_{x^2-y^2}({\bf R}) \sim (A_>-B_>)\exp(-|{\bf R}|/\xi_{x^2-y^2})$.  
Now if $A_>$ and $B_>$ do not reflect 
the symmetry of the ordered state we should expect
$P_{x^2-y^2}({\bf R})$ to be typically small and of indeterminate sign.  
Extracting $\xi_{x^2-y^2}$ via Eq. 5 is essentially impossible.  In our calculation,
$\langle O_+\rangle$ and $\langle O_-\rangle$ are both positive, giving exactly this problem.
This difficulty should be expected if there is anything in a calculation that prevents the system from being
in the long range ordered state.  Eq. 6 is generally a poor choice for a correlation function
outside of the ordered phase since it can be anomalously small for reasons unrelated to the strength of the correlations.
\begin{figure}[htb]
\includegraphics[width=3.0in,clip]{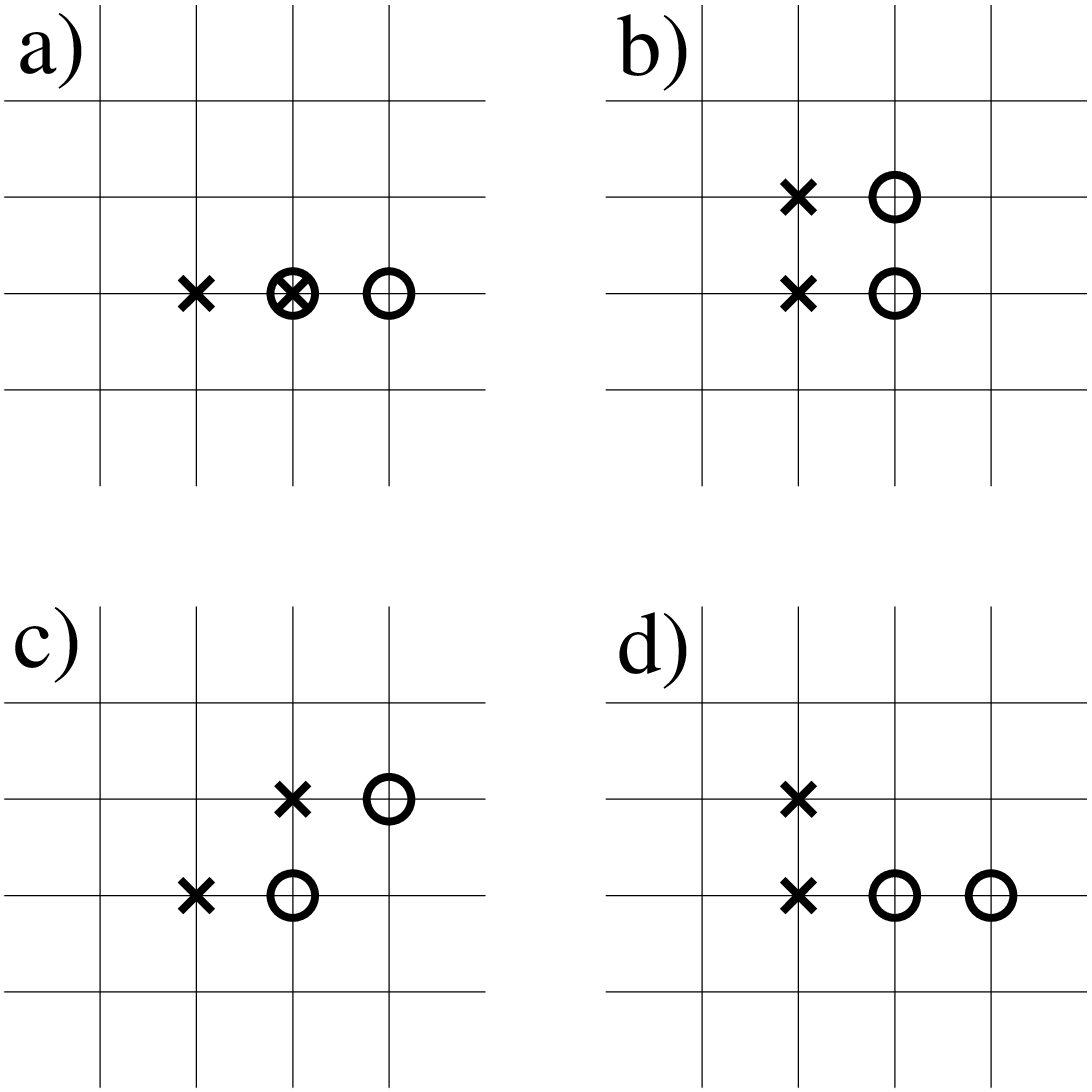}
\caption{Examples of configurations of two pairs on a square lattice.  The first pair
is represented by $\times$'s and the second by o's.  a), b) and c) are examples of contributions
to the {\bf R}=($1$, $0$) correlator.
For $s$-symmetry a), b) and c) all contribute, while for
$d_{x^2-y^2}$-symmetry the nodes eliminate c) and only a) and b) contribute.  For 
$d_{xy}$-symmetry the nodes eliminate a) and b) so only c)
contributes.  d) is an example of a possible pair correlation not included in our
definition of $P_i({\bf R})$, Eq. 7.}
\end{figure}

To reduce the possibility of spurious cancellations in the definition of
$P_i({\bf R})$ we choose to limit the $P({\bf r}, {\bf r}', {\bf R})$ we include
to having ${\bf r}={\bf r}'$.  By doing this we hold the internal degrees of freedom
of the pairs fixed while we measure the distance dependence {\bf R} of the correlation
function.
We have one pairing operator with a definite internal configuration {\bf r}
at position zero and a second pairing operator with the same internal configuration at
position {\bf R}.  We then sum over {\bf r} since we are only interested in the {\bf R} dependence.
Calculating a correlator where we allow {\bf r} and ${\bf r}'$ to vary
independently mixes distance correlation with internal degrees of freedom, which isn't the
quantity we want to measure.
Our definition of $P_i({\bf R})$ is
\begin{equation}
P_i({\bf R}) = \sum_{\bf r}\phi_i^2({\bf r}) P({\bf r}, {\bf r}, {\bf R}).
\end{equation}
With this definition of $P_i({\bf R})$ we are guaranteed to not have cancellation in
the {\bf r} sum due to trivial symmetry reasons.
The effects of different symmetries are still present through the nodes of the pair
wave function where $\phi_i({\bf r})=0$.  The correlator defined in Eq. 7 is more versatile than
the correlator defined by Eq. 6 due to Eq. 7 being able to find
growing fluctuations in the disordered phase and also show the order paramenter in the ordered phase.
An important feature of Eq. 7
is that to distinguish $s$ and $d_{x^2-y^2}$ pairing we need
to include pairs with $|{\bf r}|>1$.  Also, to have $\xi^2_{xy}$
positive we need to include pairs with $|{\bf r}|>1$.

With $P_i({\bf R})$
defined by Eq. 7, series can be calculated for $\xi_i$ defined by Eq. 5.
The resulting series are extrapolated
to low temperatures using Pad\'e approximants\cite{guttmann} either in the original expansion variable
$\beta J$ or in the transformed variable $w=\tanh(\alpha\beta J)$ 
used by Singh and Glenister\cite{singh}.  Here $\alpha$ is a numerical factor which
can be adjusted to improve the convergence of the Pad\'es.
The estimated errors are from the spread of the Pad\'e approximants at low temperatures.
\begin{figure}[htb]
\includegraphics[width=3.4in,clip]{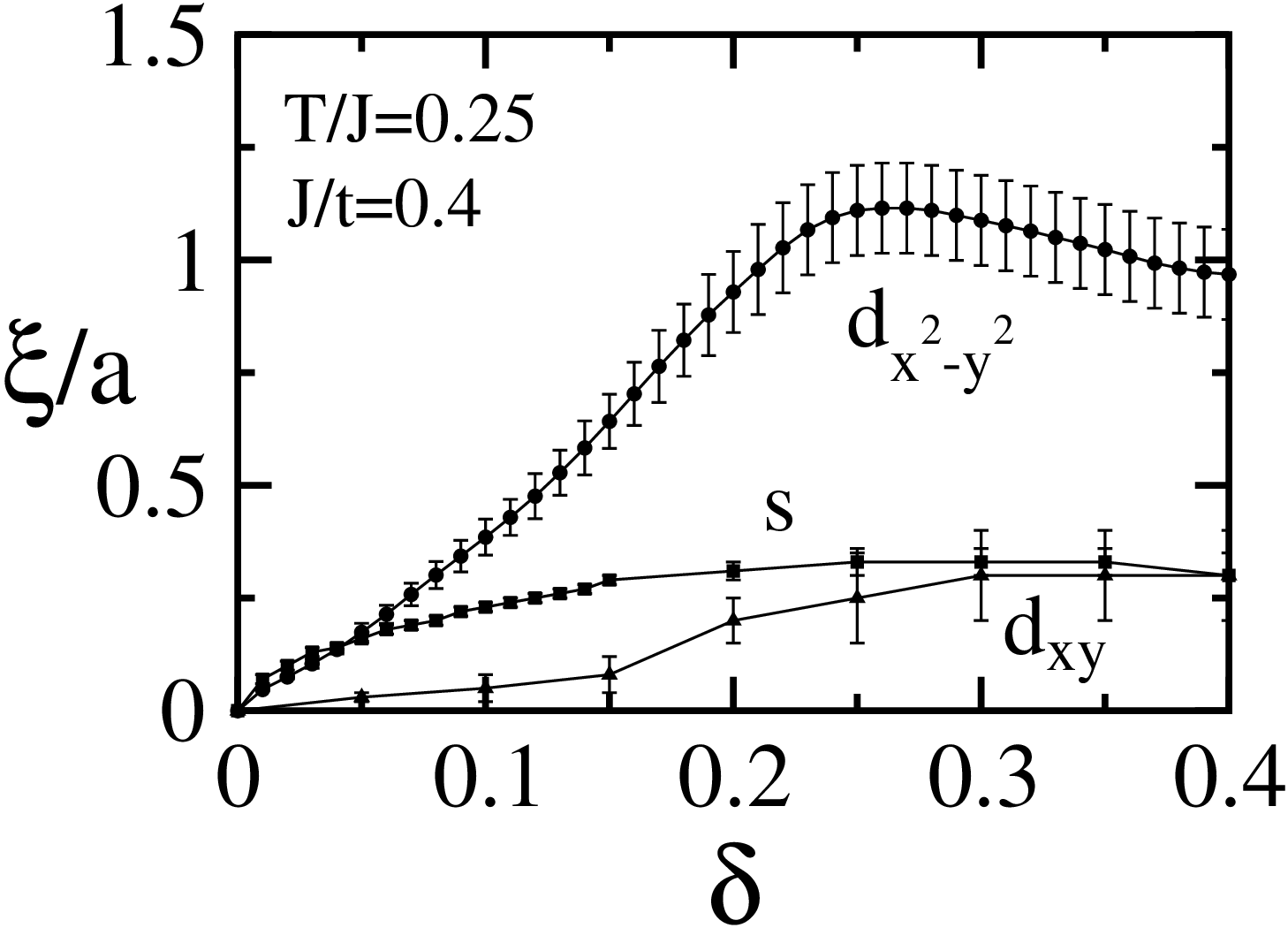}
\caption{Pairing correlation length versus doping.  The $t$-$J$ model parameter ratio is
$J/t=0.4$ and the temperature is $T/J=0.25$.  Results for $s$-, $d_{x^2-y^2}$- and
$d_{xy}$-symmetry pairing are shown.  The lines are guides for the eye.}
\end{figure}

Our results at $T/J=0.25$ are shown in Fig. 3.
Clearly the 2D $t$-$J$ model does not have $s$- or $d_{xy}$-symmetry pairing for the range
of parameters shown.  However, the correlation length for $d_{x^2-y^2}$-symmetry pairing
$\xi_{x^2-y^2}$
is large and suggests the ground state of the 2D $t$-$J$ model has long range superconducting
order with $d_{x^2-y^2}$ symmetry.  Further support for this interpretation is shown in
Fig. 4, where $\xi_{x^2-y^2}$ versus doping is plotted for a range
of temperatures.
The temperature dependence shows $\xi_{x^2-y^2}$ grows rapidly with decreasing
temperature over most of the doping range shown in Fig. 4, though
$\xi_{x^2-y^2}$ decreases with decreasing temperature for $\delta\ltrsim0.04$
This is distinct from $s$- and $d_{xy}$-symmetry pairing, which for low temperatures have 
$\xi_s$ and $\xi_{xy}$ slowly decreasing
with decreasing temperature over the whole doping range shown.
The temperature scale for growth 
in $\xi_{x^2-y^2}$
also agrees well with the temperature scales found for two-point
correlation functions\cite{putikka}.
\begin{figure}[htb]
\includegraphics[width=3.4in,clip]{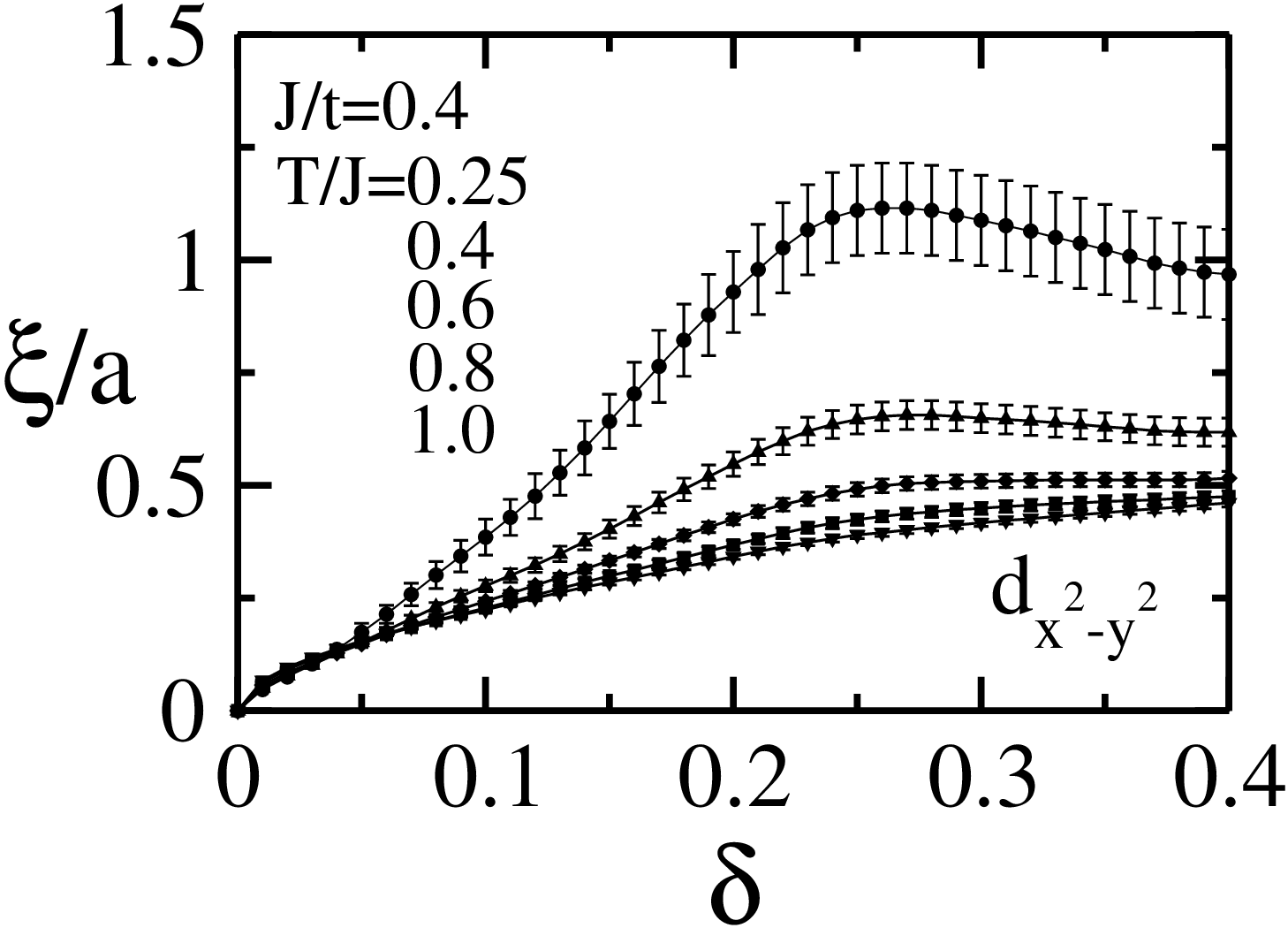}
\caption{Pairing correlation length for $d_{x^2-y^2}$-symmetry versus doping for a range
of temperatures.  The $t$-$J$ model parameter ratio is $J/t=0.4$.  The data sets
are in the order indicated on the plot, with the highest curve corresponding to 
$T/J=0.25$.  The lines are guides for the eye.}
\end{figure}

Our results are in accord with experiments on high temperature 
superconductors\cite{ginzberg}.
The pairing has $d_{x^2-y^2}$ symmetry, with the doping range where $\xi_{x^2-y^2}$ is large
comparable to where $T_{\rm c}$ is observed.  The maximum in $\xi_{x^2-y^2}$ is shifting
to smaller doping as the temperature is reduced, and likely is near $\delta\approx0.2$
for $T/J\approx0.1$, though calculations at lower temperatures are needed to show this. 
Also, $\xi_{x^2-y^2}$ grows very
steeply on the underdoped side, while falling off more slowly for overdoping.  This is due to
$\xi=0$ at half filling for all symmetries and all temperatures and the strong
antiferromagnetic correlations present for underdoping.  
In experiments $T_{\rm c}$
frequently jumps from zero to a non-zero value for underdoped samples with a small change in
doping.  

Band structure estimates\cite{pavarini} for $t$-$J$ model parameters give a next nearest
neighbor hopping term $t'$ roughly the same size as $J$.  This term has not been included
in the calculations presented here.  We expect that $t'$ would allow holes to move more
easily through the antiferromagnetic background present in the $t$-$J$ model, and thus would
move the peak in $\xi_{x^2-y^2}$ to lower doping.  Although a $t'$ term is not necessary to
have strong superconducting correlations, 
calculations including a $t'$ term should improve detailed
comparisons between experiments on high temperature superconductors and the $t$-$J$ model.

The 2D superconducting correlation length does not need to grow to an
extreme size like $\xi_{AF}$ at half filling\cite{chak} 
to produce true 3D
long range order due to the effective interplanar hopping $t_{\perp}$ driving the 2D to 3D crossover for
superconductors being
much larger than the interplanar
spin coupling $J_{\perp}$.  The temperature dependence for $\xi_{x^2-y^2}$ shown in Fig. 4 is likely to
produce $T_{\rm c}$'s in the range observed for high temperature superconductors.
Lower temperature results are needed to show this conclusively.

A peak in  $\xi_{x^2-y^2}$ has not been
observed in previous calculations.
The most likely reason is the problem with cancellation in $P_i({\bf R})$ discussed
above.  Previous calculations\cite{threv, hirsch, white, imada, scalapino, zhang,
shih, sorella, pryadko, ogata}
the authors have examined have mixed terms for anisotropic pairing correlators.
The negative $U$ Hubbard model does not have this
problem due to the pairing being predominantly on-site and isotropic\cite{moreo}.  
The ${\bf q}=0$ pairing
susceptibility was calculated by high temperature series in Ref. \onlinecite{pryadko}, 
with the conclusion
that $d_{x^2-y^2}$ pairing is not found in the 2D $t$-$J$ model.
The results of Ref. \onlinecite{pryadko} are influenced by single
particle terms we have removed and the cancellation effect.  We believe these
effects distort the results of Ref. \onlinecite{pryadko} and when properly accounted
for the ${\bf q}=0$ pairing susceptibility will show $d_{x^2-y^2}$ correlations
are present in the 2D $t$-$J$ model.

In conclusion, we find strong $d_{x^2-y^2}$-symmetry pairing correlations in the 2D
$t$-$J$ model for physical values of the model parameters and doping.  This suggests
the origin of high temperature superconductivity is the strong electronic correlations found in
the CuO$_2$ planes.

This calculation was started while W. O. P. was visiting ETH, supported by the Swiss
National Science Foundation and by a faculty travel grant from the Office of International
Studies at The Ohio State University, and
continued while taking part in a workshop on high temperature superconductors at KITP, supported
by NSF grant PHY99-07949.  W. O. P. was also partially supported by NSF grant DMR01-05659.  M. U. L. was
supported by EPSRC Grant No. GR/L86852.  The computations were done on the Itanium II cluster 
at the Ohio Supercomputer Center.  We thank Tin-Lun Ho, Ciriyam Jayaprakash, Mohit Randeria,
Maurice Rice, Doug Scalapino, Rajiv Singh, David Stroud and John Wilkins for many useful
discussions.

{\it Note added in proof}: After submission of this paper we received Ref. \onlinecite{ogata}
where calculations related to ours are reported.


\begin{references}
\bibitem{bednorz} J. G. Bednorz and K. A. M\"uller, Z. Phys. B{\bf 64}, 189 (1986).

\bibitem{anderson} P. W. Anderson, Science {\bf 235}, 1196 (1987).

\bibitem{zhangrice} F. C. Zhang and T. M. Rice, Phys. Rev. B{\bf 37}, 3759 (1988).

\bibitem{threv} For reviews see D. J. Scalapino, Phys. Rep. {\bf 250}, 329 (1995) and
E. Dagotto, Rev. Mod. Phys. {\bf 66}, 763 (1994).

\bibitem{hirsch} J. E. Hirsch and H. Q. Lin, Phys. Rev. B{\bf 37}, 5070 (1988).

\bibitem{white} S. R. White, {\it et al.}, Phys. Rev. B{\bf 39}, 839 (1989); S. R. White,
{\it et al.}, Phys. Rev. B{\bf 40}, 506 (1989).

\bibitem{imada} M. Imada, J. Phys. Soc. Jpn. {\bf 60}, 2740 (1991).

\bibitem{scalapino} D. J. Scalapino, S. R. White and S. C. Zhang, Phys. Rev. Lett.
{\bf 68}, 2830 (1992).

\bibitem{zhang} S. Zhang, J. Carlson and J. E. Gubernatis, Phys. Rev. Lett. {\bf 78},
4486 (1997).

\bibitem{shih} C. T. Shih, Y. C. Chen, H. Q. Lin and T. K. Lee, Phys. Rev. Lett. {\bf 81}, 1294 (1998).

\bibitem{sorella} S. Sorella, {\it et al.}, Phys. Rev. Lett. {\bf 88}, 117002 (2002).

\bibitem{pryadko} L. P. Pryadko, S. A. Kivelson and O. Zachar, Phys. Rev. Lett. {\bf 92},
067002 (2004).

\bibitem{gelfand} M. P. Gelfand, R. R. P. Singh and D. Huse, J. Stat. Phys. {\bf 59}, 1093 (1990).

\bibitem{conv} This example comes from conversations the authors had with T.-L. Ho, C. Jayaprakash
and M. Randeria.

\bibitem{guttmann} A. J. Guttmann, in {\it Phase Transitions and Critical Phenomena},
ed. by C. Domb and J. L. Lebowitz (Academic, San Diego, 1989) Vol. 13, p. 1.

\bibitem{singh} R. R. P. Singh and R. L. Glenister, Phys. Rev. B{\bf 46}, 11871 (1992).

\bibitem{putikka} W. O. Putikka, M. U. Luchini and R. R. P. Singh, Phys. Rev. Lett.
{\bf 81}, 2966 (1998); cond-mat/9912269.

\bibitem{ginzberg} {\it Physical Properties of High Temperature Superconductors},
Vol 1-5, ed. D. M. Ginzberg (World Scientific, Singapore).

\bibitem{chak} S. Chakravarty, B. I. Halperin and D. R. Nelson, Phys. Rev. Lett.
{\bf 60}, 1057 (1988); Phys. Rev. B{\bf 39}, 2344 (1989).

\bibitem{moreo} A. Moreo and D. J. Scalapino, Phys. Rev. Lett. {\bf 66}, 946 (1991);
A. Moreo, Phys. Rev. B{\bf 45}, 5059 (1992).

\bibitem{pavarini} E. Pavarini, I. Dasgupta, T. Saha-Dasgupta, O. Jepsen and O. K. Andersen,
Phys. Rev. Lett. {\bf 87}, 047003 (2001).

\bibitem{ogata} T. Koretsune and M. Ogata, J. Phys. Soc. Japan {\bf 74}, 1390 (2005).
\end{references}
\end{document}